\documentclass[slac_one]{revtex4}
\usepackage{graphicx}
\usepackage{fancyhdr}
\usepackage{subfigure}

\RequirePackage{xspace}
\def\invpb {\ensuremath{\mbox{\,pb}^{-1}}\xspace}
\def\invfb   {\ensuremath{\mbox{\,fb}^{-1}}\xspace}
\newcommand{\met}{\mbox{${E\!\!\!\!/_T}$}}
\def\pt         {\mbox{$p_T$}\xspace}
\def\Et         {\mbox{$E_T$}\xspace}
\newcommand{\gev}{\ensuremath{\mathrm{\,Ge\kern -0.1em V}}\xspace}
\newcommand{\gevc}{\ensuremath{{\mathrm{\,Ge\kern -0.1em V\!/}c}}\xspace}
\newcommand{\gevcc}{\ensuremath{{\mathrm{\,Ge\kern -0.1em V\!/}c^2}}\xspace}
\newcommand{\tev}{\ensuremath{\mathrm{\,Te\kern -0.1em V}}\xspace}

\begin{document}

\title{General Searches for New Physics}

%

\author{A.~Soha (on behalf of the CDF and H1 Collaborations)}
\affiliation{University of California, Department of Physics, One Shields Avenue, Davis, California 95616, USA}

\begin{abstract}
A model-independent global search for new physics has been performed at the
CDF experiment.  This search examines nearly $400$ final states, looking for
discrepancies between the observed data and the standard model expectation 
in populations, kinematic shapes, and the tails of the summed transverse
momentum distributions.  A new approach also searches in approximately $5000$
mass variables looking for `bumps' that may indicate resonant production 
of new particles.  The results of this global search for new physics in
$2 \invfb$ are presented.  In addition, a model-independent search for
deviations from the Standard Model prediction is performed in $e^+p$ and
$e^-p$ collisions at HERA~II using all H1 data recorded during the second
running phase.  This corresponds to integrated luminosities of
$178 \invpb$ and $159 \invpb$ for $e^+p$ and $e^-p$ collisions,
respectively.  A statistical algorithm is used to search for deviations
in the distributions of the scalar sum of transverse momenta or invariant
mass of final state particles, and to quantify their significance.
\end{abstract}

\maketitle

\thispagestyle{fancy}


\section{INTRODUCTION}

In stark contrast to most searches for new physics, which optimize for
a particular model or signature, the general searches presented here
are model-independent and include many final state particle combinations
in an effort to be highly inclusive.  The CDF global search results will
be presented first, followed by the H1 general search results.


\section{GLOBAL SEARCH AT CDF}

The model-independent global search for new high-$\pt$ physics in
$p \overline{p}$ collisions at the Tevatron using CDF has three
components~\cite{cdf_two_prd}:
{\sc Vista} examines populations and kinematic features of the high-$\pt$
data; the {\sc Bump Hunter}~\cite{cdf_thesis} searches for resonances in
invariant mass combinations; and {\sc Sleuth} looks for excesses at high
sum-$\pt$ ($\Sigma \pt$).

The CDF results use data corresponding to a luminosity of $2 \invfb$,
acquired through inclusive high-$\pt$ electron, muon, photon, and jet triggers.
Standard criteria are imposed to identify electrons, muons, taus, photons,
jets, $b$-jets, and missing transverse energy ($\met$), all with
thresholds equivalent to $\pt > 17 \gevc$.  Events are further selected
to meet offline requirements such as $\Et(e) > 25 \gev$,
$\pt(\mu) > 25 \gevc$, or $\Et(\gamma) > 60 \gev$~\cite{cdf_one_prd}.
Approximately $4.3$ million events are partitioned into $399$ exclusive
final states, and new categories are created as needed.

The strategy is to use Monte Carlo event generators such as {\sc Pythia}
and {\sc MadEvent} to represent the Standard Model (SM), and to pass the
resulting events through a GEANT-based simulation of the CDF detector
response.  The simulation is then used in a global fit to the
CDF data, to extract $43$ corrections factors.  The fit is performed
simultaneously to all final states and is subjected to external constraints.
The correction factors, which include corrections to leading order
theory cross sections, object reconstruction efficiencies, and
mis-identification rates, are then used to improve the SM prediction.
The three components ({\sc Vista}, {\sc Bump Hunter}, and {\sc Sleuth}) of
the global comparison between the data and SM prediction are performed, and
the procedure is iterated by feeding information back into the simulation
and correction factors until there is either a clear case for new physics
or all discrepancies have known sources.

\subsection{Population and Kinematic Distribution Results}
\label{sec:vista}

\begin{figure}
     \centering
     \subfigure{
          \label{fig:cdf_vista_pop}
          \includegraphics[width=.32\textwidth]{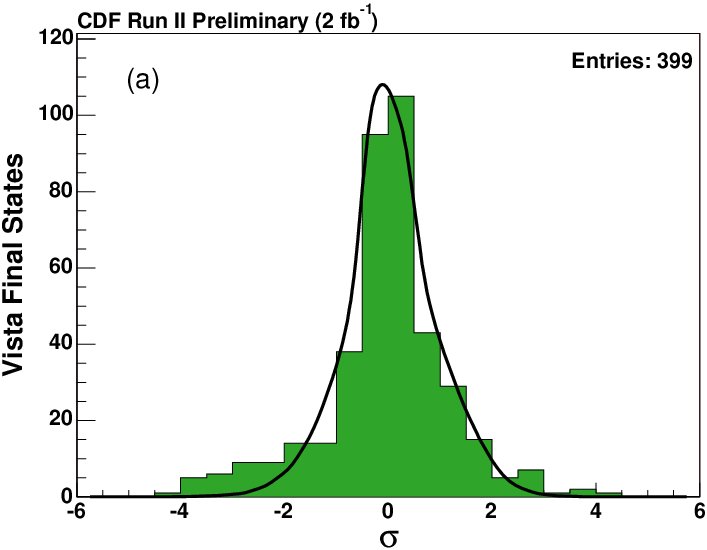}}
     \subfigure{
          \label{fig:cdf_vista_shapes}
          \includegraphics[width=.32\textwidth]{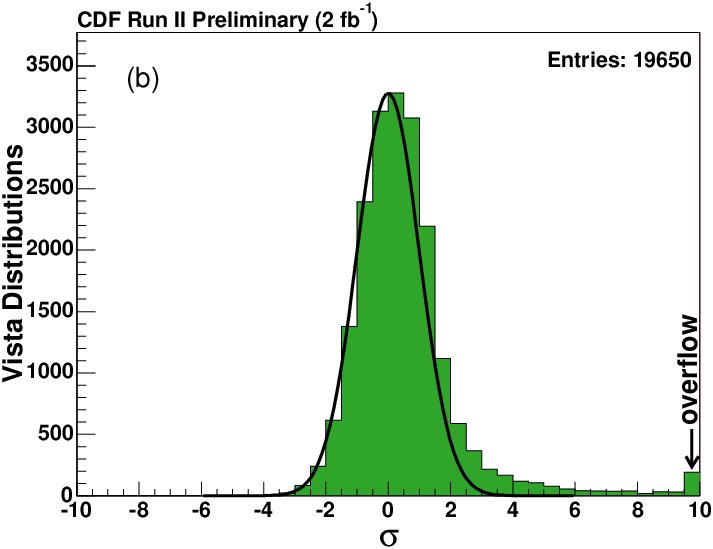}}
     \subfigure{
          \label{fig:cdf_bump_hunter}
          \includegraphics[width=.32\textwidth]{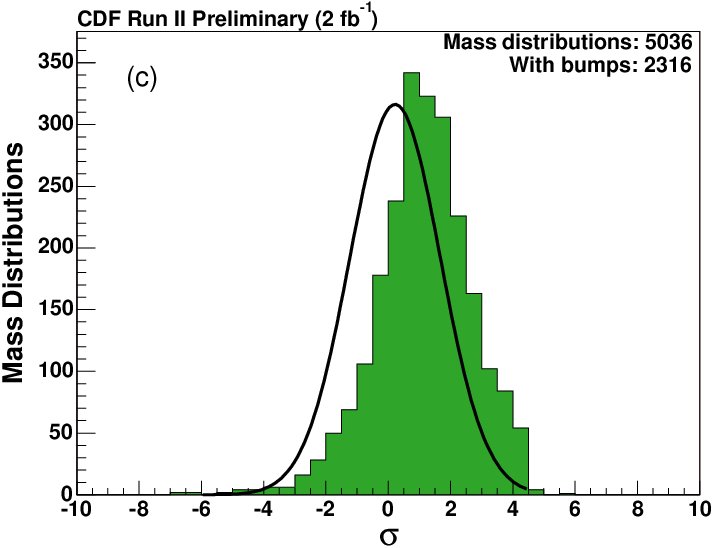}}\\
     \caption{Results for the CDF (a) population and (b) kinematic shape
comparisons, as well as the (c) bump hunter, as described in the text.}
     \label{fig:cdf_vista_and_bump}
\end{figure}

The {\sc Vista} comparison of final state populations between data and
SM predictions is shown in Fig.~\ref{fig:cdf_vista_pop}.  The histogram
shows the Poisson probability that the SM population in a final state would
fluctuate above or below the observed population in data, expressed in units
of standard deviations.  The plotted probabilities do not include
a trials factor, which accounts for the large number of final states that
are examined and reduces the significance of each observed discrepancy.
The greatest observed discrepancy is in the final state $be^{\pm}\met$
where $817.7 \pm 9.2$ events are expected and $690$ events are observed,
for a discrepancy of $-4.3 \sigma$ before the trials factor and $-2.7 \sigma$
after including the trials factor.  Therefore, no population shows a
significant discrepancy.

{\sc Vista} also automatically produces and examines $19650$ kinematic
distributions.  The results are summarized in Fig.~\ref{fig:cdf_vista_shapes},
where the histogram shows the Kilmogorov-Smirnov probability that the
distributions in the data and SM prediction are consistent, expressed in
units of standard deviations.  The trials factor due to examining thousands
of distributions has not yet been accounted for in the plot.  Distributions
are considered discrepant if they disagree by more than $5\sigma$
(approximately $>3\sigma$ after including the trials factor).  The
$555$ distributions that meet this criteria are examined more closely.
It turns out that $81\%$ of the discrepancies can be explained by a
deficiency in modeling soft jet emission in QCD parton showering.  An
additional $16\%$ are due to inadequate modeling of the transverse boost
of the colliding system and $3\%$ are due to residual crudeness in the
correction factor model, mostly from using simplified $\pt$-dependencies
in fake rate correction factors.  Therefore, there are no claims
for new physics based on the kinematic distribution comparisons.

\subsection{Bump Hunter Results}

A new resonance might appear as a bump in an invariant mass distribution.
The CDF {\sc Bump Hunter} uses the final states from {\sc Vista} to form
all invariant mass combinations and perform a comparison between data and
SM backgrounds.  A search window of $2 \Delta M$, where $\Delta M$ is the
expected detector mass resolution, is scanned across each invariant mass
distribution.  A candidate bump must have at least five data events
and side-bands that are in better agreement than the central search window.
Pseudo-experiments are then used to estimate the significance of any
qualifying bumps.  The results are shown in Fig.~\ref{fig:cdf_bump_hunter},
which shows the probability for a corresponding bump from pseudo-data to
have a larger significance than the one found in data, cast in terms of
standard deviations.  Of the $5036$ scanned distributions, $2316$ have
qualifying bumps.  The visible shift in the histogram is caused by local
deficiencies in the SM prediction, but does not invalidate the method
since the shift makes it more likely that a bump surpasses the threshold for
further study.  The threshold for further investigation is $5\sigma$, which
corresponds to $3 \sigma$ after including the trials factor for $5036$ mass
distributions.  There is one bump beyond this threshold, in a final state
with four jets and low $\Sigma \pt$, but it is found to be due to the same
soft jet modeling problem mentioned in section~\ref{sec:vista}.  Hence,
no new physics is found by the {\sc Bump Hunter} in $2 \invfb$.

\subsection{Search at High Sum-\boldmath{\pt}}

{\sc Sleuth} assumes that new physics will appear as an excess, and that
the excess will be at high $\Sigma \pt$ and in one final state.
The $\Sigma \pt$ is the scalar sum of the $\pt$ of the individual objects,
unclustered energy, and $\met$.  For each final state, the $\Sigma \pt$
distribution is scanned, and the one-sided region with the most significant
excess of data is selected.  The significance, $\mathcal{P}$, is determined
as the fraction of pseudo-experiments that find a region at least as
discrepant as the one observed in data.  The final state with the largest
discrepancy is $e^\pm\mu^\pm$, with $\mathcal{P}=0.00055$, corresponding to
$3.26\sigma$ before including a trials factor.  It is found that $8\%$ of
experiments like CDF would find an excess at least as large as this most
discrepant {\sc Sleuth} final state.  There are no claims of new physics
using {\sc Sleuth} with $2 \invfb$.


\section{GENERAL SEARCH AT H1}

The global search for new physics at H1 uses data corresponding to
luminosities of $178 \invpb$ and $159 \invpb$ from $e^+p$ and $e^-p$
collisions, respectively~\cite{H1_update,H1_117}.  Isolated electrons,
muons, photons, jets, and
neutrinos are included if they have $\pt > 20 \gevc$ and a polar angle
satisfying $10^\circ < \theta < 140^\circ$.  The event selection requires
exclusive final states with two or more objects, and events are classified by
the number and types of objects.  This procedure examines all combinations
and finds that $23$ final states are populated.  Simulations are used for
all contributing SM processes, including photoproduction, deep-inelastic
scattering, QED Compton scattering, electroweak production, and QCD.
The resulting predictions are used in comparisons to the event yields,
$\Sigma \pt$, and invariant mass distributions found in data.  A statistical
algorithm is then employed to identify the largest deviations and evaluate
the associated probabilities.

\subsection{Event Yields, Sum-\boldmath{\pt}, and Invariant Mass}

\begin{figure}
     \centering
     \subfigure{
          \label{fig:H1_pop_a}
          \includegraphics[width=.48\textwidth]{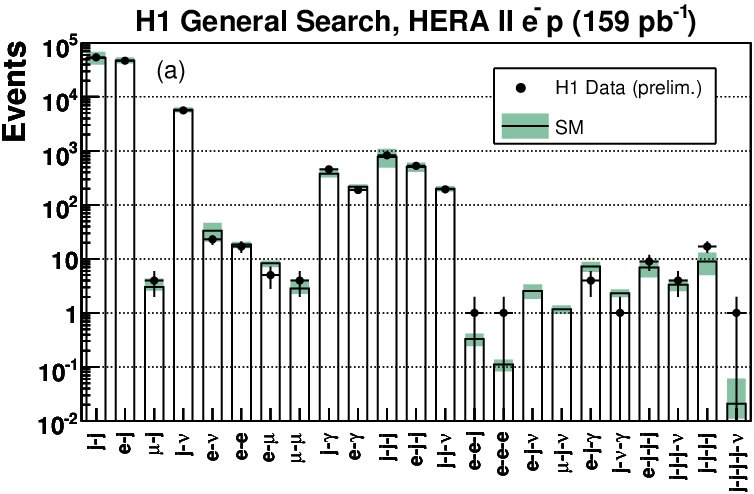}}
     \subfigure{
          \label{fig:H1_pop_b}
          \includegraphics[width=.48\textwidth]{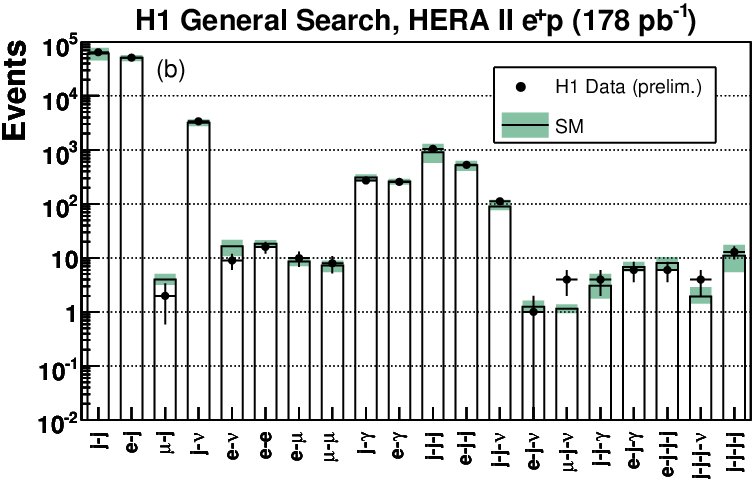}}\\
     \caption{Event yield comparisons between data and SM predictions
using (a) $e^-p$ and (b) $e^+p$ collisions.  The plots include categories
with $\ge 1$ data event or $> 1$ expected SM events.  Error bars include
theory uncertainty and experimental systematics.}
     \label{fig:H1_pop}
\end{figure}

\begin{figure}
     \centering
     \subfigure{
          \label{fig:H1_mAll_minus}
          \includegraphics[width=.48\textwidth]{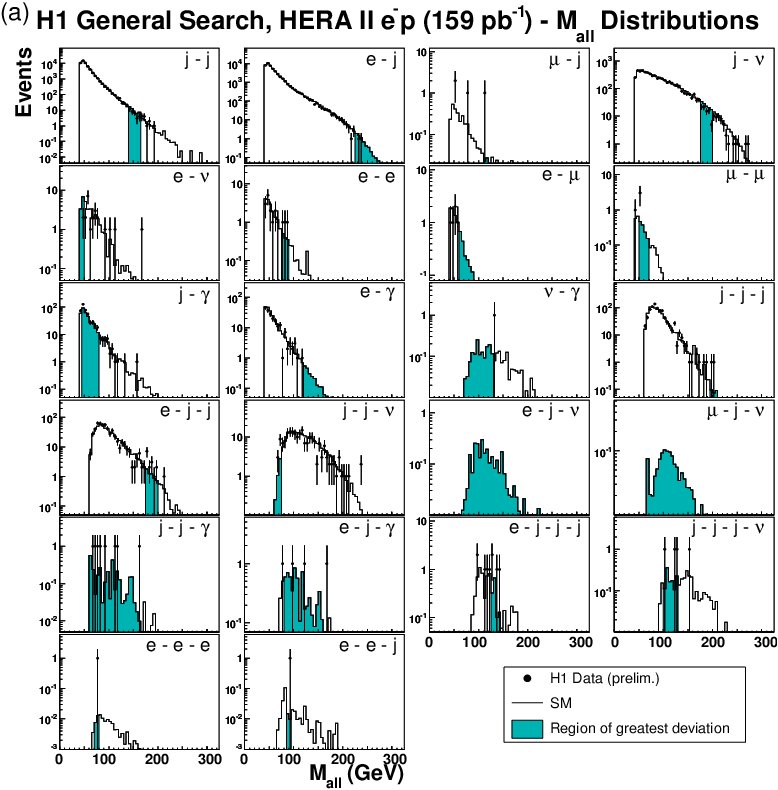}}
     \subfigure{
          \label{fig:H1_sumpt_plus}
          \includegraphics[width=.48\textwidth]{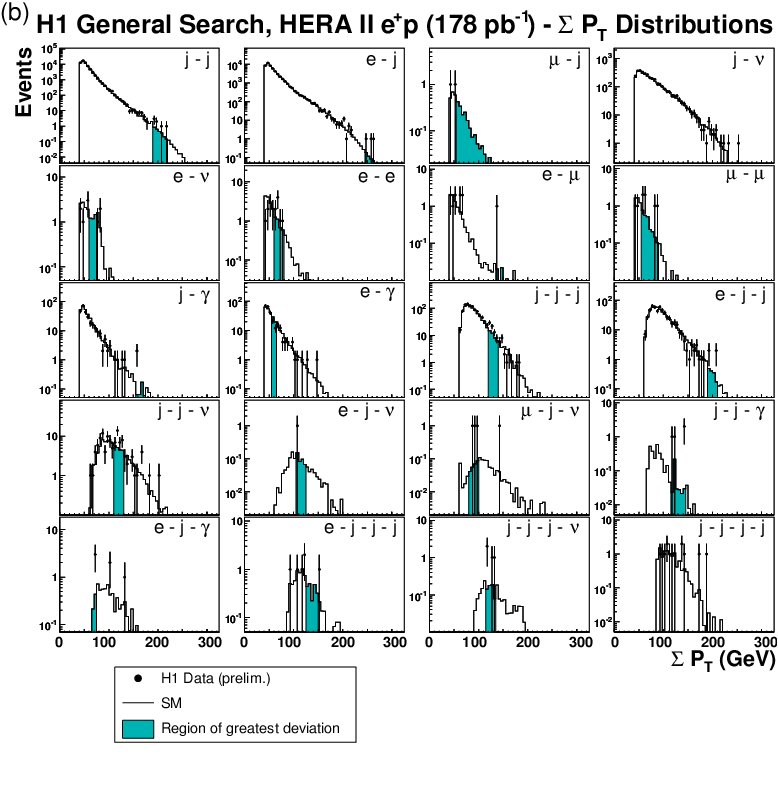}}\\
     \caption{Distributions of data events and SM expectations
for the (a) invariant mass in $e^-p$ collisions and (b) $\Sigma \pt$ in
$e^+p$ collisions.  The regions of greatest deviation are shaded.}
     \label{fig:H1_sigmapt}
\end{figure}

The event yield comparisons show good agreement in all final states,
in both the $e^-p$ and $e^+p$ data, as shown in Fig.~\ref{fig:H1_pop}.
The $\Sigma \pt$ and invariant mass distribution comparisons are conducted
by finding regions of greatest deviation between data and the SM expectation
for each final state.  All groups of neighboring $5 \gev$ bins are tested.
A measure, p, is determined as the probability for a positive or negative
fluctuation of the SM expectation to be at least as large as that observed
in data.  The procedure accounts for Poisson statistical errors and Gaussian
systematic uncertainties.  For each final state, the region with the smallest
p-value is selected.  The results for the invariant mass comparisons using
$e^-p$ data are shown in Fig.~\ref{fig:H1_mAll_minus} and the results for
the $\Sigma \pt$ comparison using $e^+p$ data are shown in
Fig.~\ref{fig:H1_sumpt_plus}.

\subsection{Significance}

\begin{figure}
     \centering
     \subfigure{
          \label{fig:H1_phat_a}
          \includegraphics[width=.48\textwidth]{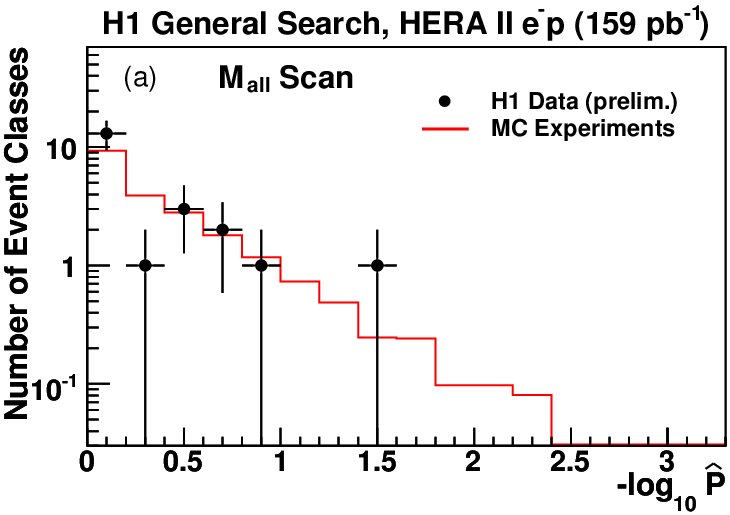}}
     \subfigure{
          \label{fig:H1_phat_b}
          \includegraphics[width=.48\textwidth]{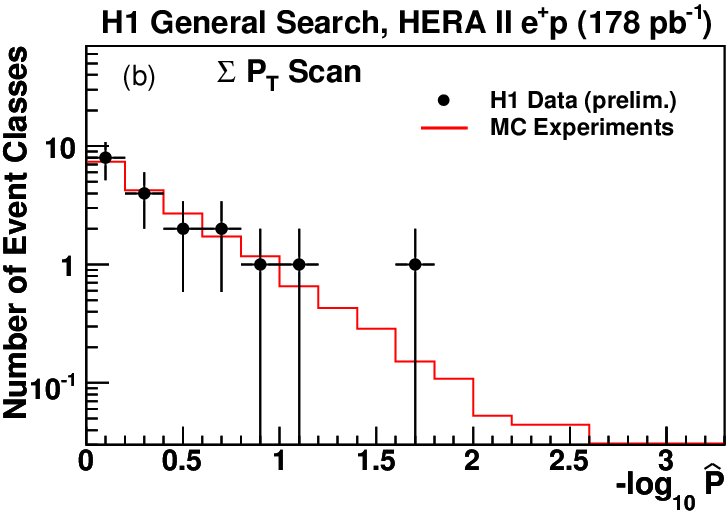}}\\
     \caption{Distributions of the quantity $-\log_{10} \hat{\rm P}$
for all of the final states used in the comparisons of the (a) invariant mass
in $e^-p$ data and (b) $\Sigma \pt$ in $e^+p$ data.  The plotted lines show
the expectation from Monte Carlo experiments.}
     \label{fig:H1_phat}
\end{figure}

The significance of each deviation in the H1 $\Sigma \pt$ and invariant mass
comparison is evaluated using pseudo-data.  The method determines the
probability, $\hat{\rm P}$, to observe a region with a p-value less than the
smallest p-value seen in data.  Calculating this $\hat{\rm P}$ allows for
the comparison of deviations across different final state categories.
Fig.~\ref{fig:H1_phat_a} shows the $-\log_{10} \hat{\rm P}$ distribution
for the invariant mass comparison using $e^-p$ data, while
Fig.~\ref{fig:H1_phat_b} shows the $-\log_{10} \hat{\rm P}$ distribution
for the $\Sigma \pt$ comparison using $e^+p$ data.
Note that a $5\sigma$ discrepancy would correspond to a value of
$-\log_{10} \hat{\rm P}$ between $5$ and $6$.  No such significant
discrepancies between data and SM expectations are observed.  The largest
deviation is in the $\mu j \nu$ final state category.

\section{CONCLUSIONS}

The CDF and H1 general searches for new physics have probed large
datasets for indications of new physics in population and kinematic
distributions, using a large number of final states.
These searches provide broad views of the high-$\pt$ data samples and
demonstrate understanding of the detectors and SM simulation.
They do not rule out all sources of new physics, thus leaving open the
possibility for future discoveries.


\begin{acknowledgments}
The author wishes to thank the 2008 ICHEP organizers and hosts, as
well as the CDF and H1 collaborations and funding sources.
\end{acknowledgments}

\end{document}